\documentclass[aps,prb,twocolumn,showpacs,floatfix]{revtex4-1}
\usepackage{graphicx}
\usepackage{color}
\usepackage{amsmath}
\usepackage{amssymb}
\usepackage{amsfonts}
\usepackage{bm}
\definecolor{scarred}{rgb}{0.75,0.0,0.0}
\usepackage{hyperref}
\hypersetup{
colorlinks=true,final=true,
        linkcolor=blue,
        citecolor=blue,
        filecolor=blue,
        urlcolor=blue,
}

\begin{document}
\title{Transient Floquet engineering of superconductivity} 
\author{Nagamalleswararao Dasari}\email{nagamalleswararao.d@gmail.com}
\author{Martin Eckstein}\email{martin.eckstein@fau.de}
\affiliation{Department of Physics, University of Erlangen-Nuremberg, 91058 Erlangen, Germany}

\begin{abstract}
Intense time-periodic laser fields can transform the electronic structure of a solid into strongly modified Floquet-Bloch bands. While this suggests multiple pathways to induce electronic orders such as superconductivity or charge density waves, the possibility of preparing low-energy phases of Floquet Hamiltonians remains unclear because of the energy absorption at typical experimentally accessible driving frequencies. Here we investigate a realistic pathway towards laser control of electronic orders, which is the transient enhancement of fluctuating orders. Using a conserving Keldysh Green's function formalism, we simulate the build-up of short range Cooper-pair correlations out of a normal metal in the driven attractive Hubbard model. Even for frequencies only slightly above or within the bandwidth, a substantial enhancement of correlations can be achieved before the system reaches a high electronic temperature.  This behavior relies on the non-thermal nature of the driven state. The effective temperature of the electrons at the Fermi surface, which more closely determines the superconducting correlations, remains lower than an estimate from the global energy density. Even though short ranged, the fluctuations can have marked signatures in the electronic spectra.
\end{abstract}

\maketitle
\section{INTRODUCTION}

The availability of ultra-short and highly intense light pulses has inspired a very fruitful experimental agenda of 
controlling the properties of quantum materials on ultra-short times.\cite{Giannetti2016,Basov2017} In this context, a 
question of both fundamental and practical interest is whether it is possible to enhance or even induce macroscopically 
coherent electronic orders in a solid. Experiments in this direction include possible light-induced superconductivity in 
cuprates\cite{fausti2011,Hu2014} or fullerides,\cite{Mitrano2016} or the strengthening of an excitonic condensate 
through photo-excitation.\cite{mor2017} 

Among the possible pathways to control material properties, ``Floquet engineering'' is particularly appealing 
from a theoretical prospective: Under a time-periodic perturbation, 
such as the electric field of a laser or a coherently excited phonon, the evolution and  steady states of a quantum system are 
described (after suitably averaging over a period) by an effective Floquet-Hamiltonian, which can be entirely different 
from the un-driven Hamiltonian.\cite{Goldman2014, Bukov2015, Eckardt2015} A simple variant of this idea is the control 
of the band structure by off-resonant laser fields. In the limit of high frequency, an oscillating electric field 
with projection $E(t)=E_0\cos(\Omega t)$ along a bond of the lattice renormalizes the tunneling matrix 
element $t_\text{hop}$ between orbitals along that bond to\cite{Dunlap1986}
\begin{align}
\label{thefirstequation}
t_\text{hop} \to t_\text{hop}\,\mathcal{J}_0(eaE_0/\hbar\Omega).
\end{align} 
 Here $\mathcal{J}_0(x)$ is the zeroth order Bessel function, $a$ is the bond length, and $e$ the electron charge. 
Furthermore, Floquet theory has been used to predict topologically nontrivial Floquet-Bloch bands,
\cite{Oka2009, Kitagawa2011, Jotzu2015,Claassen2017} and possibilities to manipulate magnetic exchange interaction 
in Mott insulators\cite{Mentink2015, Goerg2018, Mikhaylovskiy2015,Claassen2016} or phonon-mediated pairing interactions.\cite{Knap2016, Kennes2017, Komnik2017} 

Already the simple high-frequency result \eqref{thefirstequation} suggests many pathways to manipulate electronic orders, for instance by changing the ratio of interaction and bandwidth, or by changing the shape of the Fermi-surface.\cite{Kennes2018} Floquet-Bloch bands have indeed been observed in solids,\cite{Wang2013} but a main hindrance towards a control of {\em low-energy} orders in the steady state is the energy absorption from the periodic drive.\cite{Murakami2017} In theoretical few-band models, one can choose off-resonant frequencies sufficiently far above the bandwidth, so that heating is slow and nontrivial Floquet pre-thermalized states\cite{Bukov2015PRL, Canovi2016, Weidinger2017} can emerge, but in real materials there will be further electronic transitions at higher energies. For the interesting case of inducing new orders out of a metallic phase it is yet unclear in general whether low-energy states of the Floquet Hamiltonian, or non-thermal driven states with non-trivial properties, can be reached in practice. 

An alternative direction for experiments will therefore be to analyze the transient build-up of electronic orders. A possible manifestation of such a transient effect has been reported in the organic charge-order material $\alpha$-(ET)$_2$I$_3$, which shows a reduction of the reflectivity, indicating stronger insulating behavior, in response to a few-cycle pulse with a frequency right above the absorption band.\cite{Ishikawa2014} (Note that already few cycle pulses can lead to a similar effective Hamiltonian as for the periodic drive.\cite{Eckstein2017}) In the present work, we demonstrate the feasibility of transient Floquet engineering in a theoretical model, and show that short-range superconducting order can be enhanced in a normal metal following the bandwidth control by a laser, even though true long range order does not form before heating sets in. 

This setting brings up another fundamental question, i.e., how, on short times, a symmetry 
broken state is born out of an initial normal (disordered) phase.  When studying dynamical symmetry breaking 
within time-dependent (dynamical) mean-field theory, one has to break the symmetry in the initial state with a 
small (usually global) order parameter, which then grows exponentially in time and non-homogeneously in 
space.\cite{Tsuji2013b, Sentef2016} Such a classical description may be qualitatively valid once quasi-macroscopic 
domains have formed, but does not address the early dynamics of short-range correlations (and neither inhomogeneous effects such as defect formation through the Kibble-Zurek mechanism,\cite{Kibble1976, Zurek1985}, or domain growth\cite{Chern2018}).  Recently, a number of theoretical works have instead investigated  the growth of order out of a disordered state:  Ref.~\onlinecite{Bauer2015} uses dynamical mean-field theory to study the antiferromagnetic susceptibility in the repulsive Hubbard model 
after a slow ramp of the interaction, and finds transient regimes with strong correlations even though the 
system later thermalizes to a normal hot-electron state. In the repulsive Hubbard model with both charge (stripe) and 
d-wave superconducting correlations, variational Monte Carlo simulations result in the intriguing observation that after a 
bandwidth-renormalization superconducting correlations can be enhanced with respect to equilibrium because the 
build-up of competing charge correlations lacks behind.\cite{Ido2017} Further, an exact diagonalization study of the short 
range pairing correlations in the extended Hubbard model after a quench to the superconducting regime finds 
optical signatures (a Drude peak) similar to experiments, even though the system is not long-range 
ordered.\cite{Bittner2017} Finally, Refs.~\onlinecite{Lemonik2017,Lemonik2018} analyze the critical dynamics of 
superconducting fluctuations close to a pairing instability.
Taking the electrons at fixed temperature, the slower evolution of the superconducting correlations shows universal behavior with intriguing experimental fingerprints in the optical conductivity and the electronic spectra. 

In the present work, we focus on the attractive Hubbard model as a paradigmatic model for 
superconducting pairing, and simulate the dynamics while the system is driven by an electric field 
with frequency $\Omega$. In the Floquet picture, superconductivity is favored in the driven state as a simple consequence of enhancing the ratio of interaction and bandwidth by the factor $1/\mathcal{J}_0$, see Eq.~\eqref{thefirstequation}. For 
experimentally accessible frequencies slightly above or within the bandwidth, we observe that a transient fluctuating 
order can emerge even when a long-range ordered state does not form. The study of short-range correlations requires a proper treatment of the momentum-dependent collective orders and their feedback on the momentum-dependent electronic self-energy, which is in general more demanding than a static or even dynamical mean-field treatment.\cite{aoki2014_rev} Our simulations build on an earlier implementation of the time-dependent GW formalism, which was used to study the melting of excitonic order in the presence of dynamic screening processes.\cite{Golez2016} 

The paper is organized as follows. In Sec.~\ref{methodssec} we define the model and explain the 
diagrammatic equations. Section.~\ref{hsvkqxaz} briefly recapitulates the equilibrium solution of 
the model. In Sec.~\ref{hsvkqxaz02} we then study the formation of Cooper-pair correlations out of a 
normal metal after a ramp-on of the interaction, which can be understood as the far off-resonant limit of Floquet theory.
 In Sec.~\ref{hsvkqxaz03} we contrast these results with 
the behavior when the ratio of interaction and bandwidth is increased by a time-periodic electric field, 
and we analyze the nature of the driven state. In Sec.~\ref{hsvkqxaz04} we investigate the effect of the transient order on the  electronic spectra of the driven state, and Sec.~\ref{hsvkqxaz001} gives a summary and conclusion.

\section{MODEL AND METHOD}
\label{methodssec}

\subsection{Model}

We study the two-dimensional attractive Hubbard model,
\begin{align}
H 
=
-\!\sum_{\langle j,l \rangle,\sigma}
\!\!t_{jl}\,
c_{j\sigma}^\dagger
c_{l\sigma}
+
U
\sum_{j}
n_{j,\uparrow}
n_{j,\downarrow}
-
\mu
\sum_{j,\sigma}
n_{j,\sigma}.
\label{THEmodel}
\end{align}
Here $c_{j\sigma}^\dagger$ creates an electron with spin $\sigma\in\{\uparrow,\downarrow\}$ on 
site $j$ of a lattice. $t_{jl}$ is a nearest neighbour hopping, and $U$ is an attractive on-site 
interaction ($U<0$). In the numerical simulations we consider a square lattice of given size $L\times L$ with 
periodic boundary conditions, and correspondingly a Brillouin zone of $L^2$ momenta.

Without external fields, we assume an isotropic nearest neighbor hopping $t_{jl}\equiv t_\text{hop}$. The electric 
field $\bm E(t)$ of the laser is incorporated using a gauge with zero scalar potential, such that 
$\bm E(t)=-\frac{1}{c}\frac{\partial \bm A(t)}{\partial t}$, where $\bm A(t)$ is the vector potential. Using 
the Peierls substitution, the hopping along a bond between sites at positions $\bm R_l$ and $\bm R_j$ is then modified to 
\begin{align}
\label{jwhdekd}
t_{jl}(t)
=
t_\text{hop}
\exp\Big(\frac{ie}{\hbar c}(\bm R_j-\bm R_l)\bm A(t)\Big).
\end{align}
Note that Eq.~\eqref{thefirstequation} is obtained as a time average of this equation for an 
oscillating electric field, for which the projection of the vector potential along the 
bond $(j,l)$ is $A(t) = \frac{E_0c}{\Omega}\cos(\Omega t)$. If not stated otherwise, we apply electric 
fields along the $(1,1)$-direction of the lattice, such that all bonds are affected in the same way, and choose 
units such that $a=1$, $c=1$, $e=1$, and $\hbar=1$; $t_{\text{hop}}=1$ sets the energy scale. In momentum space, the dispersion is $\epsilon(\bm k) \equiv \epsilon_0(\bm k)$ = -2$t_\text{hop}[\cos(k_x a) + \cos(k_y a)]$ in 
zero field, and  $\epsilon(\bm k,t) = \epsilon_0(\bm k-\bm A(t))$ otherwise. 

\subsection{Formalism}

\begin{figure}[tbp]
\includegraphics{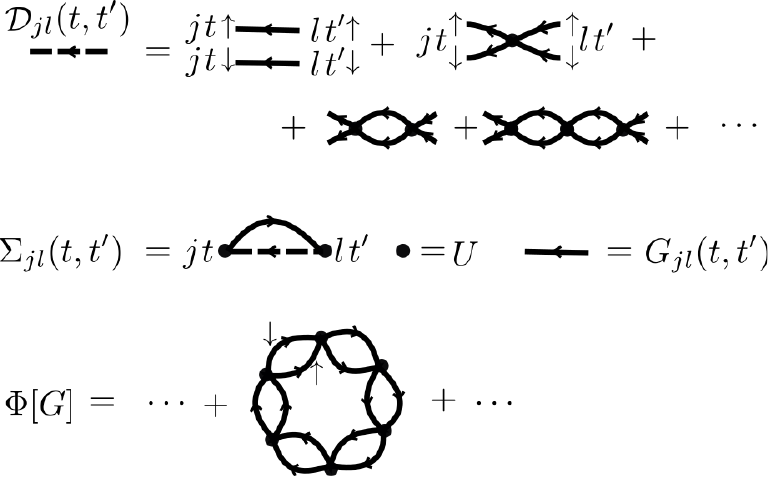}
\caption{Feynman diagrams for the Cooper-pair fluctuations $\mathcal{D}_{jl}(t,t')$ (bold dashed line) and the 
self-energy $\Sigma_{jl}(t,t')$ in terms of the fully interacting Green's functions (bold line). The 
last line shows an exemplary contribution to the Luttinger-Ward functional.}
\label{Fig:Diagrams}
\end{figure}

To study the non-equilibrium dynamics of the model, we employ the Keldysh formalism on the $L$-shaped time
 contour $\mathcal{C}$, which allows to describe the unitary dynamics of an isolated quantum system 
starting from an initial equilibrium state at given temperature $T$. (For an introduction to the 
formalism and the notation, see, e.g., Ref.~\onlinecite{aoki2014_rev}). Diagrammatic approaches 
developed for finite temperature equilibrium states can be directly rewritten within the Keldysh 
formalism. For this work, we use a formalism designed to describe the interplay of electrons and 
pairing fluctuations in the normal state. We introduce the contour-ordered electronic Green's functions
\begin{align}
G_{jl}(t,t') 
=
-i 
\langle
T_{\mathcal{C}} c_{j}(t) c_{l}^\dagger (t')
\rangle,
\end{align}
and the propagator for the Cooper-pair fluctuations
\begin{align}
\label{kqxbkl}
\mathcal{D}_{jl}(t,t') 
=
-i 
\langle
T_{\mathcal{C}} \hat \Delta_{l}(t) \hat \Delta_{l}^\dagger (t')
\rangle,
\end{align}
where $ \hat \Delta_{j} = c_{j\uparrow} c_{j\downarrow}$.
Equations of motion for these propagators in the normal phase are obtained by using a 
Hubbard-Stratonovich decoupling of the interaction in the pairing channel, treating the dynamics 
of the field in the saddle-point approximation around $\langle\hat\Delta\rangle=0$. On the diagrammatic 
level, this corresponds to expanding $\mathcal{D}$ in particle-particle ladder diagrams of the electronic 
Green's function. The electronic self-energy $\Sigma$ then includes the interaction of the electrons 
with the field $\mathcal{D}$ (see Fig.~\ref{Fig:Diagrams}). The formalism sums the subset of diagrams of the 
fluctuation-exchange interaction\cite{Bickers1989} which are most relevant for the superconducting instability. 
It has been discussed in equilibrium,\cite{Deisz1998, Engelbrecht2000} in particular to 
investigate fingerprints of normal-state superconducting fluctuations on the electronic spectrum. We therefore 
only briefly summarize the equations in Sec.~\eqref{kbqxjka} below. 

We use a self-consistent formulation of the diagrammatic equations, i.e., $\mathcal{D}$ and $\Sigma$ are 
expanded in terms of the fully interacting Green's function. The self-energy functional $\Sigma[G]$ is then 
derivable from a Luttinger-Ward functional $\Phi[G]$, $\Sigma_{jl}(t,t')= \delta \Phi[G]/\delta G_{lj}(t',t)$ 
(see Fig.~\ref{Fig:Diagrams}). This ensures energy and particle number conservation,\cite{Baym1961} which is 
particularly important to study the evolution of the total energy in the non-equilibrium dynamics. Interestingly it 
has also been found that the self-consistent expansion qualitatively well captures the normal state behavior of the 
correlation length in the two-dimensional system, which undergoes a Berezinsky-Kosterlitz-Thouless 
transition \cite{Engelbrecht2000} (see also Sec.~\ref{hsvkqxaz}).

The attractive Hubbard model has also a sub-leading instability towards charge density wave order, which becomes 
degenerate with superconductivity at half filling. While there are predictions to manipulate the relative strength 
of the two orders in non-equilibrium, both by time-dependent protocols\cite{Sentef2017} and by electric currents\cite{Matthies2018}, the 
present formalism captures only the superconducting instability. To study the competition of short-range transient Cooper-pair and charge 
density wave correlations would definitely be interesting, but requires a different diagrammatic approach, which is left for future work.

\subsection{Implementation}
\label{kbqxjka}

In real-space, the diagrammatic equations depicted in Fig.~\ref{Fig:Diagrams} read as follows. The pairing correlations satisfy the integral equation
\begin{align}
\label{jsjhw;qxk}
\mathcal{D}_{jl}(t,t')
=
\mathcal{D}^0_{jl}(t,t')
+
\sum_m
\int_\mathcal{C}\!\!
d\bar t\,\,
\mathcal{D}^0_{jm}(t,\bar t)
U(\bar t) 
\mathcal{D}_{ml}(\bar t,t'),
\end{align}
where 
$\mathcal{D}^0_{jl}= iG_{jl}(t,t')G_{jl}(t,t')$ is the bare pairing correlation. The electronic self-energy is then given by 
\begin{align}
\label{cdblacsx}
\Sigma_{jl}(t,t')
=
-i V_{jl}(t,t') G_{lj}(t',t),
\end{align}
with $V_{jl}(t,t')=U(t) \mathcal{D}_{jl}(t,t') U(t')$. 
All real-space functions depend only on space difference, and we solve the equations in momentum space on a finite momentum grid.  
After Fourier transform, Eq.~\eqref{jsjhw;qxk} becomes
\begin{align}
\label{lwkndcwkqx}
\mathcal{D}_{\bm q}(t,t')
=
\mathcal{D}^{0}_{\bm q}(t,t')
+
\int_\mathcal{C}\! d\bar t\,\,
\mathcal{D}^{0}_{\bm q}(t,\bar t)
U(\bar t)
\mathcal{D}_{\bm q}(\bar t,t'),
\end{align}
with 
\begin{align}
\label{kqcebqxl02}
\mathcal{D}^{0}_{\bm q} (t,t') = \frac{i}{L^2} \sum_{\bm k} G_{\bm k}(t,t') G_{\bm q-\bm k}(t,t').
\end{align}
The self-energy \eqref{cdblacsx} in momentum space is given by
\begin{align}
\label{kqcebqxl01}
\Sigma_{\bm k} (t,t') = \frac{-i}{L^2} \sum_{\bm q} V_{\bm q}(t,t') G_{\bm q-\bm k}(t',t),
\end{align}
with $V_{\bm q}(t,t')= U(t)  \mathcal{D}_{\bm q}(t,t') U(t')$. Finally, the momentum-dependent electronic 
Green's functions satisfy the Dyson equation
\begin{align}
\label{dyson-g}
G_{\bm k} (t,t') = [i\partial_t - \epsilon(\bm k,t) - \Sigma_H(t) - \Sigma_{\bm k}(t,t')]^{-1},
\end{align}
with the Hartree self-energy $\Sigma_H(t) = U \langle n(t)\rangle$.

The numerical solution of Eqs.~\eqref{lwkndcwkqx} to \eqref{dyson-g} is performed on a 
finite momentum grid of $L\times L$ points in the Brillouin zone. The number of independent $\bm k$-points 
depends on the symmetry of the problem. The latter is reduced in the presence of an external field, which is why 
simulations for nonzero field will be performed for smaller lattices. Equations \eqref{dyson-g} and \eqref{lwkndcwkqx} are 
integral equations on $\mathcal{C}$, which can be solved using high-order accurate algorithms for Volterra integral 
equations.\cite{aoki2014_rev,Eckstein2010} The main numerical bottleneck is the memory required to store the double-time 
functions $G_{\bm k}$ and $\mathcal{D}_{\bm k}$ at each $\bm k$. Equations \eqref{dyson-g} and \eqref{lwkndcwkqx} can be 
parallelized on several computing nodes, but the evaluation of the momentum sums in \eqref{kqcebqxl01} and \eqref{kqcebqxl02} 
then requires a collective communication.  

\subsection{Observables}

In Sec.~\ref{lsxlxa} below we analyze in particular the behavior of the pairing correlations in real space, which 
are obtained from the function \eqref{kqxbkl} at equal time,
\begin{align}
\label{kjbxqkxjbq}
D(\bm R_j-\bm R_l,t)
&\equiv
\langle \hat \Delta_{l}^\dagger (t) \hat \Delta_{j}(t) 
\rangle
=
i
\mathcal{D}^<_{jl}(t,t)
\\
&=
\frac{1}{L^2}\sum_{\bm q} e^{i\bm q(\bm R_j-\bm R_l)}
i\mathcal{D}_{\bm q}^<(t,t).	
\end{align}
Another important observable to be discussed is the total energy density, $E_{tot}(t)=\langle H(t)\rangle$. We have
\begin{align}
E_{tot}(t)
&=
E_{kin}+E_{int},
\label{etot.snx}
\\
E_{kin}(t)
&=
\frac{1}{L^2}
\sum_{\bm k,\sigma}
\epsilon(\bm k,t)
n_{\bm k,\sigma}(t)
\\
\label{xsnlqn;k}
E_{int}(t)
&=
\frac{1}{L^2}
\sum_{\bm k}
(-i)[\Sigma_{\bm k}\ast G_{\bm k}]^<(t,t)
\end{align}
where $n_{\bm k,\sigma}(t)=-iG_{\bm k}^<(t,t)$. In the second equation, the symbol $\ast$ denotes the convolution along $\mathcal{C}$, and Eq.~\eqref{xsnlqn;k} is obtained from the equation of motion for $G$. In the numerical implementation, we have confirmed that $E_{tot}(t)$ remains time-independent up to the numerical accuracy when $H(t)$ is time-independent (e.g., after an interaction quench), which must be the case because a conserving approximation is used for the self-energy.

\section{RESULTS}
\label{lsxlxa}

\subsection{Equilibrium properties}
\label{hsvkqxaz}

\begin{figure}[tbp]
\begin{center}
\includegraphics[width=0.9\columnwidth]{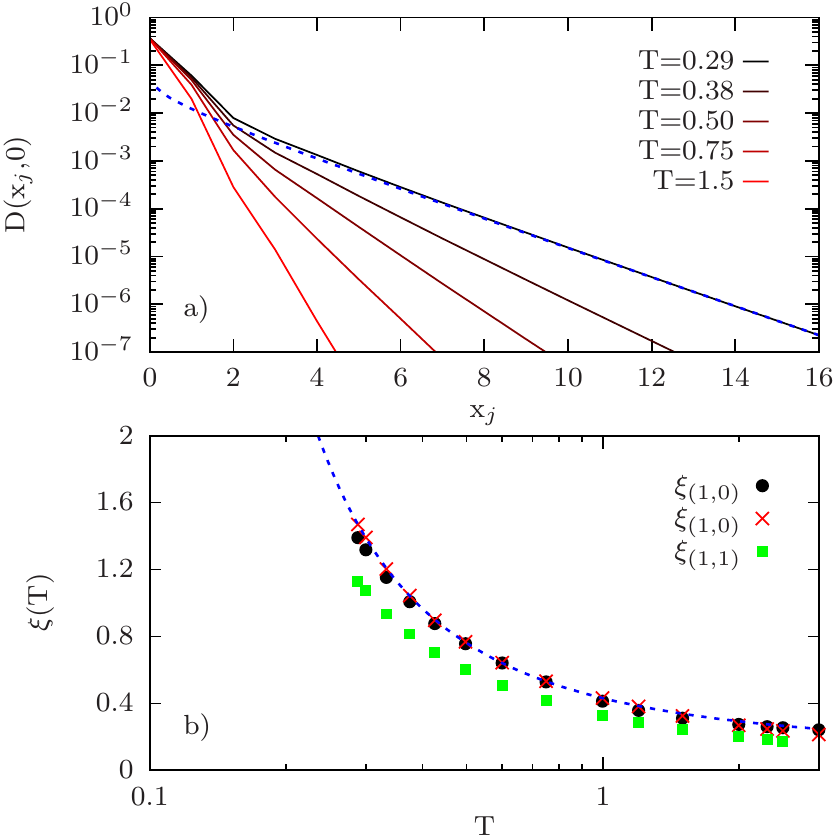}
\caption{a) Pairing correlations $D(x_j,0)$ along the $(1,0)$ direction of 
the lattice for interaction $U=-3$ and various temperatures $T$. 
$L=70$ is the lattice size. The dashed line shows a fit with Eq.~\eqref{e,bslxs;} in 
the range $10\le x_j \le 30$. b) Correlation length $\xi(T)$ extracted from the fit of the 
data in panel a) with Eq.~\eqref{e,bslxs;} in the range $10\le x_j \le 30$ (cross symbols) 
and $4\le x_j \le 8$ (circles). The line plots Eq.~\eqref{e,bslxs;01} with $A=1.275$ 
and $T_\text{BKT}=0.03$. Square symbols show the correlation length extracted from $D(x_j,x_j) $ along the $(1,1)$-direction.}
\label{fig:fig1aaa}
\end{center}
\end{figure}

Before studying the driven system, we briefly summarize the equilibrium properties of the model. 
Figure \ref{fig:fig1aaa}a shows the pairing correlations $D(\bm R)$ [Eq.~\eqref{kjbxqkxjbq}] as a function of 
distance along the (1,0)-direction of the lattice, $D(x_j,0)$. For all temperatures, we observe a decay at 
large distances, with an increase of the correlation length with decreasing temperature. In two 
dimensions, true long range order is not possible for $T>0$, but the system is expected to undergo a 
Berezinsky-Kosterlitz-Thouless (BKT) transition at some temperature $T_\text{BKT}$. In the normal phase, 
correlations asymptotically decay like 
\begin{align}
\label{e,bslxs;}
D(|\bm R|) \sim \frac{1}{|\bm R|^{1/4}}e^{-|\bm R|/\xi(T)},
\end{align}
with a temperature-dependent correlation length $\xi(T)$ which diverges at the BKT transition like
\begin{align}
\label{e,bslxs;01}
\xi(T)\sim \exp(A/\sqrt{T-T_\text{BKT}}).
\end{align}
We extract $\xi(T)$ from a fit to the numerical data in Fig.~\ref{fig:fig1aaa}a with Eq.~\eqref{e,bslxs;}. The 
correlation length is shown in Fig.~\ref{fig:fig1aaa}b, together with a fit (blue dashed curve) 
representing Eq.~\eqref{e,bslxs;01}. Because the temperatures accessed in this investigation are 
considerably larger than $T_\text{BKT}$, fitting Eq.~\eqref{e,bslxs;01} does not give a very accurate value for $T_\text{BKT}$, 
although it has been noted that the critical region in the two-dimensional Hubbard model is relatively 
wide.\cite{Engelbrecht2000} In order to reach lower temperatures, one would have to study larger 
system sizes $L$ to ensure that $L\gg \xi(T)$. For the present analysis this turns out to be not necessary, 
because the correlation length reached in the driven states remains of the same order as in Fig.~\ref{fig:fig1aaa}.

For later reference we also note that in the present regime the correlation length can be estimated already 
accurately from relatively short distances $4\le x_j \le 8$, see the black circles in Fig.~\ref{fig:fig1aaa}b. 
Furthermore, although $\xi$ is only of the order of few lattice constants, the decay of correlations in space is 
already fairly isotropic, and the correlation lengths extracted along the $(1,0)$ and $(1,1)$-directions do not 
differ much (compare cross and square symbols in Fig.~\ref{fig:fig1aaa}b).

\subsection{Interaction ramp}
\label{hsvkqxaz02}

\begin{figure}[tbp]
\begin{center}
\includegraphics[width=0.9\columnwidth]{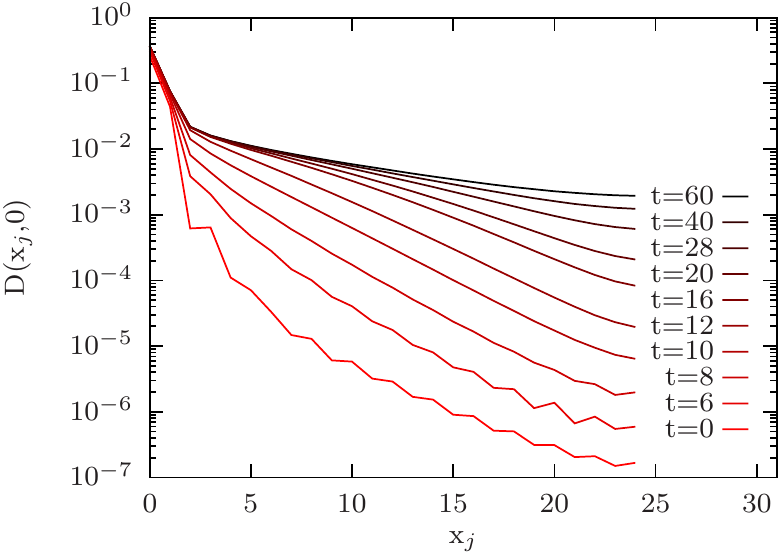}
\caption{Pairing correlations $D(\bm R)$ along the $(1,0)$-direction ($\bm R=(x_j,0)$) for a
ramp of interaction from $U_{i}=-0.5$ to $U_f=-3.0$, plotted at various times. 
(Lattice size $L=50$, initial temperature $T=1/15$).}
\label{fig:fig1}
\end{center}
\end{figure}

Before analyzing the field-driven systems we study the build-up of pairing correlations in the 
Hubbard model after an artificial increase of the interaction. In this way the energy can 
change only during the ramp, and a controlled renormalization of the ratio $|U|$/bandwidth 
is obtained without energy absorption from a drive. We ramp the interaction between 
values $U_i$ and $U_f$ according to the protocol
\[
  U(t)= 
\begin{cases}
    U_i+(U_f-U_i)\sin(\pi t / 2 t_c)^2 & \text{for } t\leq t_c\\
    U_f              & \text{for } t> t_c\\
\end{cases}.
\]
The ramp duration $t_c=12$ is chosen slow enough such that the system remains close to adiabatic. 
(For a sudden quench ($t_c=0$) the system is strongly excited, so that pairing correlations simply decay with time).

In Fig.~\ref{fig:fig1}, we plot the pairing correlations $D(\bm R)$ along the (1,0)-direction of the square 
lattice for a ramp $U_{i}=-0.5$ to $U_f=-3.0$. In the initial state the Cooper-pair correlations decay on the 
scale of few lattice sites. They start to grow in space  during and  after the pulse, and finally approach a steady regime. 
At intermediate times the behaviour of $D(\bm R)$ signals the existence of two length scales. For example, 
the curve at $t=28$ has different slopes for $j\lesssim 15$ and $j\gtrsim 15$. This may be a signature of 
spreading correlations:\cite{Cheneau2012} For large distance, the correlations still maintain a fast decay like 
in the initial state, and a new correlation length can only be established in a range $|\bm R|<v_D t$ where $v_D$ is 
some maximal speed for the spread of the correlations. Estimating the velocity of the spreading of correlations from the 
point in space where $D$ takes a given value, e.g., $D( \hat x v_D t,t)=10^{-3}$, gives $v_D\approx t_{hop}$ at 
times $t$=12 around the end of the ramp. 
Although for a detailed systematic analysis the system size is not large enough, one can see that, as expected, the correlations spread slower than the electron velocity (the maximal group velocity of electrons, at $\bm k=(\pi/2,\pi/2)$, is $v_{max}=4t_{hop}$), but fast enough to extend over the system sizes studied below within the simulated time.

At long times, the correlations saturate at values much larger than in the initial state. This is because the 
effective temperature of the system after the slow ramp is comparable to the initial temperature of the 
system (see also next section), such that even after a thermalization of the system the correlation would 
not further decay. A behavior as observed in the DMFT simulations for the anti-ferromagnetic order,\cite{Bauer2015} 
where after the ramp the system undergoes a transient regime with increased correlation length before thermalization, is 
found for shorter ramp time. In this case however, the system is strongly excited, and also the transient increase of the 
correlation length is relatively small. 

\begin{figure}[tbp]
\begin{center}
\includegraphics[width=0.9\columnwidth]{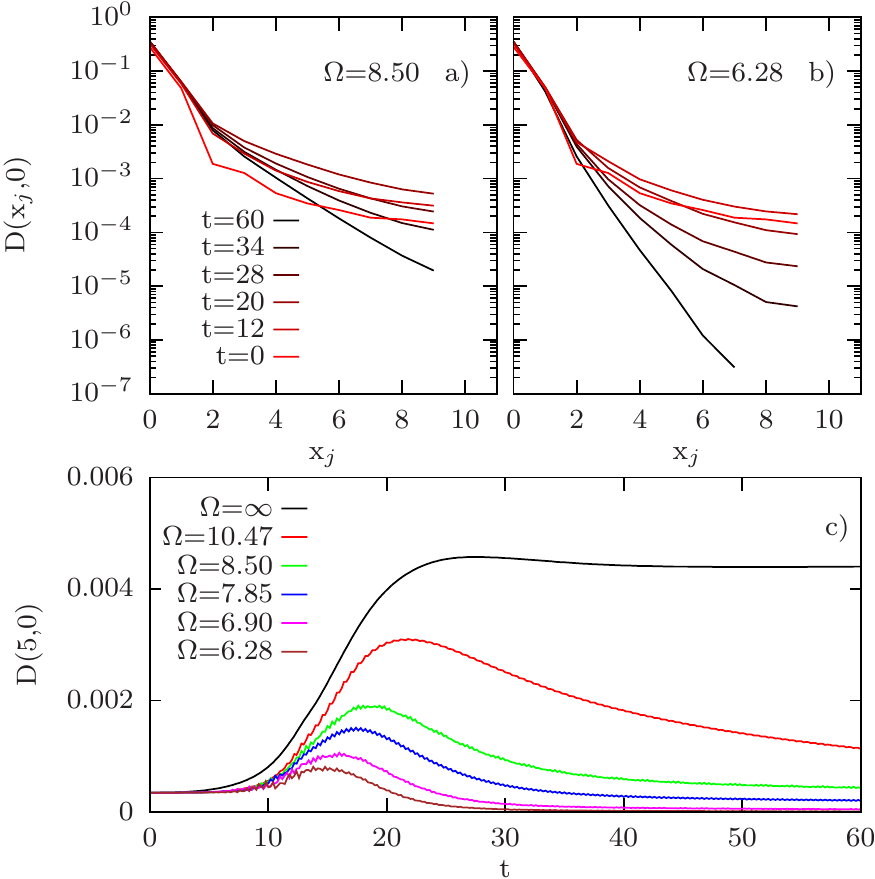}
\caption{Pairing correlations $D(\bm R)$ along the $(1,0)$-direction for various times in the field-driven 
system. The driving frequency is $\Omega=8.50$ for (a) and $\Omega=6.28$ for (b). Both simulations are 
for $U/t_{hop}=-1$, and $U/t_\text{hop}^\text{eff}=-3$ after ramp-on of the time-periodic driving. 
(Lattice size $L=20$, initial temperature $1/15$.) c) $D(\bm R)$ at $\bm R=(5,0)$ as a function of time, 
for various driving frequencies $\Omega$.}
\label{fig:fig2khqwxb}
\end{center}
\end{figure}

\subsection{Floquet band renormalization}
\label{hsvkqxaz03}

We now proceed to analyze the dynamics induced by an oscillating electric field. We choose the 
vector potential in Eq.~\eqref{jwhdekd} with projection $A(t)= A_0(t)\cos(\Omega t)$ along (1,1)-direction. A 
smaller lattice size ($L=20$) is consider for the field-driven simulations, 
because the reduced lattice symmetries in the driven case requires more 
momentum points (see Sec.~\ref{methodssec}). Within a time $t_c$ the amplitude $A_0(t)$ is ramped up to a 
final vale $A_f$ (correspondingly the amplitude of the electric field is $E_0=A_f\Omega$), with a ramp profile $A_0(t)=(t/t_c)^2A_f$ for $t<t_c$ and $A_0(t)=A_f$ for $t>t_c$. 
We choose $A_f=1.8114$, such that 
after the ramp the effective hopping is reduced by a factor $t_\text{hop}^\text{eff} /t_{\text{hop}}= \mathcal{J}_0(A_f)=1/3$ 
[c.f.~Eq.~\eqref{thefirstequation}], and the ratio $|U|$/bandwidth is increased by a factor three. In the following, we will 
refer to the attractive Hubbard \eqref{THEmodel} model with $U=-1$ and $t_\text{hop}=1/3$ simply as the ``effective Hamiltonian'', and 
compare the properties of the driven system with the equilibrium properties of the latter.

Figure~\ref{fig:fig2khqwxb}a and b show the emerging pairing correlations for various times and driving 
frequencies  $\Omega=8.50$ and $\Omega=6.28$ slightly above and below the non-interacting bandwidth $W=8t_\text{hop}$. 
For the earlier times, we observe an increase of the correlations similar to the behavior after the 
interaction ramp [Fig.~\ref{fig:fig1}], but i the driven system pair correlations steadily decrease at later times. The non-monotonous evolution is illustrated by the time-dependence of $D(\bm R)$ at a given point $\bm R=(5,0)$, see Fig.~\ref{fig:fig2khqwxb}c. Only for $\Omega=\infty$, which is simulated as a time-dependent 
ramp of the hopping amplitude given by $t_\text{hop}^\text{eff}(t)=t_\text{hop} \mathcal{J}_0(A_0(t))$, one observes an 
increase which prevails towards long times. The analysis shows that with realistic pulses 
frequencies close to the bandwidth one can achieve a significant enhancement of the superconducting 
correlations, in spite of the energy absorption which, as we will see now, is the reason for the suppression of the order 
at longer times.

\begin{figure}[tbp]
\begin{center}
\includegraphics[width=0.9\columnwidth]{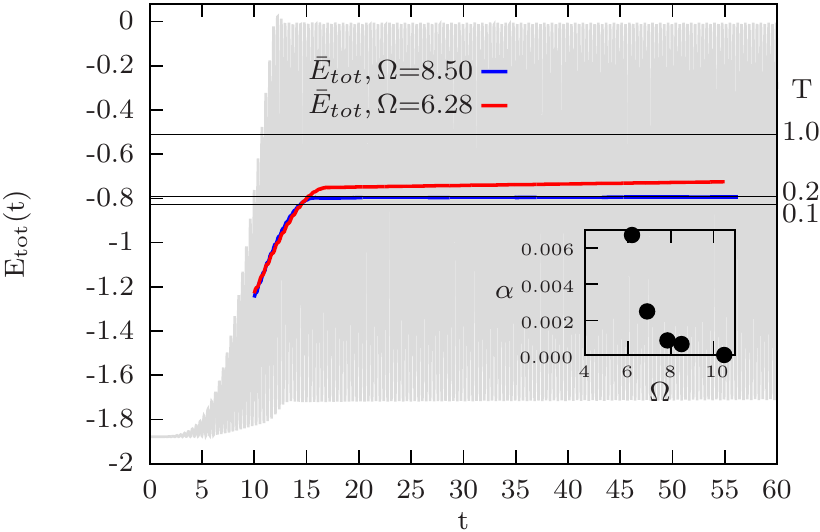}
\caption{Time-dependent energy $E_{tot}(t)$ of the driven system [c.f.~Eq.~\eqref{etot.snx}]: The black and red lines give 
the moving average of the energy over $10$ periods, for driving frequencies $\Omega=8.50$ and $6.28$ within and outside the bandwidth. The full 
oscillations of $E_{tot}$ are indicated by the light grey lines for $\Omega=10.47$; the amplitude of the oscillations is similar in 
magnitude for other frequencies. The labels at the right vertical axis show the temperature of an un-driven system in equilibrium 
with renormalized hopping $t_\text{hop}=1/3$ and an energy density corresponding to the left vertical axis. Inset: Energy absorption of the system for different driving frequencies after the ramp.}
\label{fig:fig2khqsmxb}
\end{center}
\end{figure}

\begin{figure*}[tbp]
\begin{center}
\includegraphics[width=1.7\columnwidth]{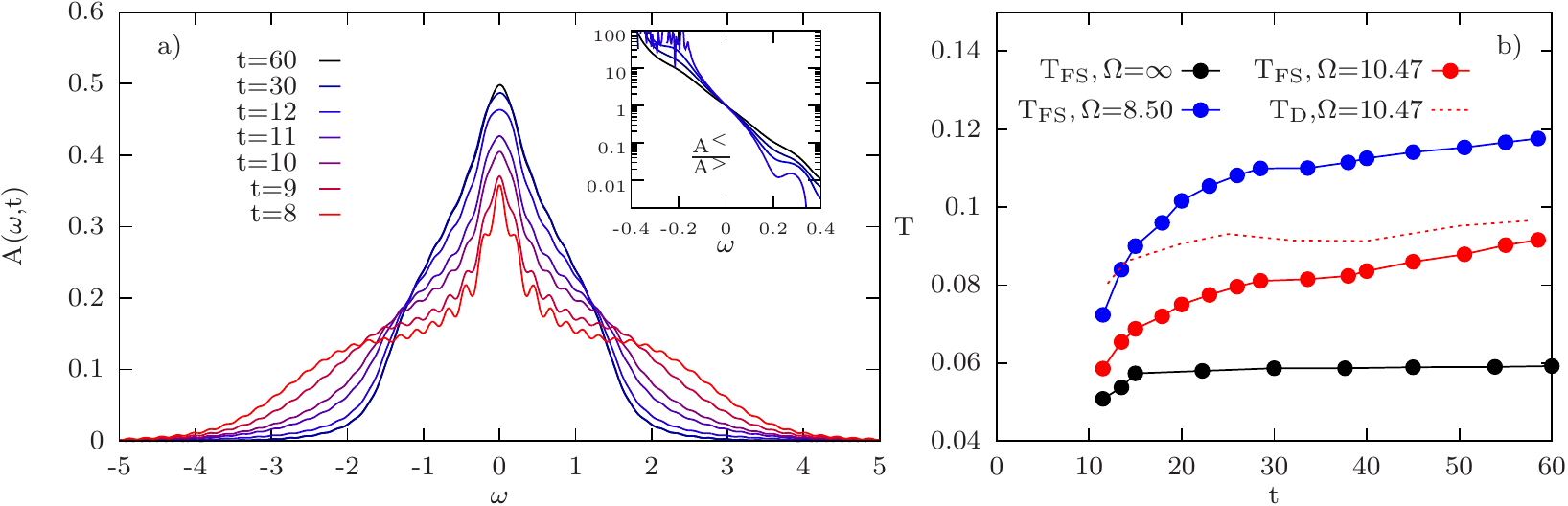}
\caption{(a) The local spectral function for a driving frequency $\Omega=10.47$ at different times. Oscillations on the spectra 
at early times are due to finite cut-off of the Fourier integral in Eq.~\eqref{eq:FT}.
Inset: The ratio $\kappa(\omega,t)=\ln\Big[\frac{A^{<}(\omega,t)}{A^{>}(\omega,t)}\Big]$ for the same driving frequency. (b) Temperature $T_\text{FS}$ 
[Eq.~\eqref{khabqcblca}] for different driving frequencies.}
\label{fig:fig3}
\end{center}
\end{figure*}

In order to test to what extent the decrease of the pairing correlations at long time is 
explained by the energy absorption, we evaluate the total energy $E_{tot}(t)$ [Eq.~\eqref{etot.snx}], 
and compare to the energy of the effective Hamiltonian at different temperatures (Fig.~\ref{fig:fig2khqsmxb}). 
It is important to note that the time-dependence of  $E_{tot}$ (light grey line in Fig.~\ref{fig:fig2khqsmxb}), 
or the instantaneous value $E_{tot}(t)$ itself tells nothing about the energy absorption. During one cycle, $E_{tot}$ falls 
well below the ground state energy of the effective Hamiltonian, so that this value cannot be related to 
an ``effective temperature'' of the latter. The energy absorption becomes instead manifest in the 
energy $\bar E_{tot}(t) = \frac{1}{\tau}\int_{t-\tau/2}^{t+\tau/2} d\bar t \,E_{tot}(\bar t)$ averaged 
over few periods, $\tau=n2\pi/\Omega$: The latter shows a linear increase  $\bar E_{tot}(t)\sim \alpha t + \text{const.}$ after 
the ramp, with a threshold-like increase of the rate $\alpha(\Omega)$ for 
frequencies $\Omega\approx 8t_\text{hop}$ around the bandwidth (inset of Fig.~\ref{fig:fig2khqsmxb}). 

It is now a natural question whether the effective temperature estimate from the 
mean energy $\bar E_{tot}$ can explain the build-up and decay of the superconducting 
correlations. Below we will see that this is not the case: The Floquet system in equilibrium with 
the same energy density $\bar E_{tot}$ as the driven system would have lower superconducting 
correlations. A different temperature estimate can be obtained from the electronic distribution 
functions. In equilibrium, the fluctuation-dissipation theorem (FDT) for fermions gives a universal ratio between the 
occupied density of states $A^<(\omega)$ and the spectrum $A^<(\omega) = f(\omega) A(\omega)$. 
Here $A(\omega) = -\frac{1}{\pi}\text{Im} G^R(\omega)$, and $A^<(\omega)=\frac{1}{2\pi i} G^<(\omega)$, 
where  $G^<(t,t') = i\langle c(t')^\dagger c(t)\rangle$ and $G^R(t,t')=-i\theta(t-t')\langle \{ c(t),c(t')^\dagger \} \rangle$ 
are the lesser and retarded propagators respectively, and $f(\omega)$ is the Fermi function. In the driven case, we evaluate 
time-dependent spectra, and similar $A^<$, as 
\begin{equation}
A(t,\omega) = -\frac{1}{\pi} {\rm{Im}} \int^{t_{cut}}_0 ds e^{i\omega s} G^R(t+s,t),
\label{eq:FT}
\end{equation}
with $t_{cut}=30$, averaged over few driving periods (for later times an analogous backward Fourier 
transform is used). 
A convenient quantity to verify the FDT is the 
logarithmic ratio $-\ln \big[A(\omega)/A^<(\omega)-1] \equiv \ln \big[A^<(\omega)/A^>(\omega)]$, which gives a 
linear function $\kappa(\omega)=-\omega/T$ in a thermal equilibrium state. In Fig.~\ref{fig:fig3}a, we plot the 
time-dependent local spectra and the ratio $\kappa(\omega,t)\equiv \ln\Big[\frac{A^{<}(\omega,t)}{A^{>}(\omega,t)}\Big]$ for $\Omega=10.47$. One can see that $\kappa(\omega,t)$ is linear around $\omega=0$, which represents the energy 
range of electrons close to the Fermi surface, while away from the Fermi surface the distribution functions take a 
more non-thermal form. We therefore extract an effective temperature $T_\text{FS}(t)$ of the electrons at the Fermi surface as
\begin{align}
\label{khabqcblca}
\frac{1}{T_\text{FS}(t)}
=
-\frac{d}{d\omega}
\ln\Big(\frac{A^{<}(\omega,t)}{A^{>}(\omega,t)}\Big)_{\omega=0},
\end{align}
This temperature turns out to be consistently lower than the temperature obtained from the 
average energy (Fig.~\ref{fig:fig3}b): For $\Omega=8.50$ and $t=60$, e.g., the energy density $\bar E_{tot}(t)$ corresponds 
to a temperature $T=0.192$ in the effective Hamiltonian system [Fig.~\ref{fig:fig2khqsmxb}], while $T_\text{FS}=0.117$. This indicates that 
the driven system is in a strongly non-thermal state, in which different energy regions are not yet thermalized, so that a 
description by a single effective temperature of the Floquet system is not possible. 

\begin{figure}[tbp]
\begin{center}
\includegraphics[width=0.9\columnwidth]{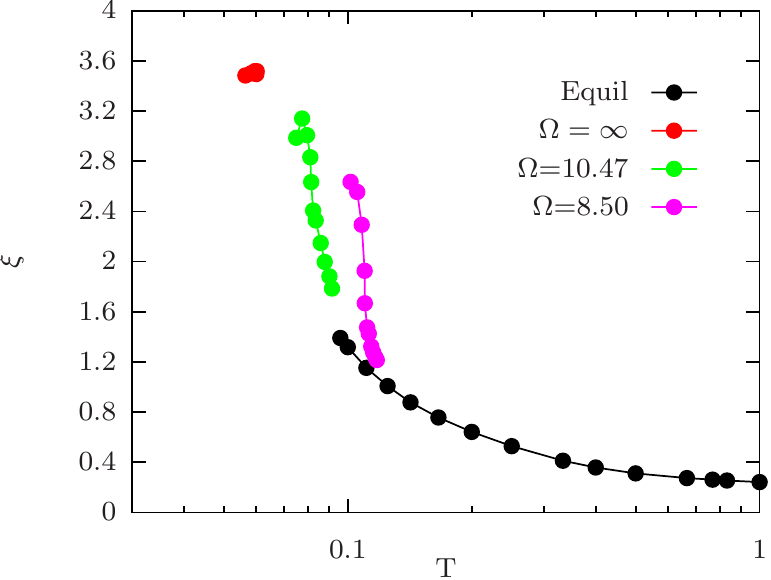}
\caption{Correlation length $\xi$ of the driven system, extracted from a fit 
with Eq.~\eqref{e,bslxs;} to the time-dependent data in the range $4\le x_j \le 8$. Black circles:  Correlation length for the effective Hamiltonian system ($U=-1$, $t_\text{hop}=-1$).}
\label{wixkwlqbjxaz}
\end{center}
\end{figure}

We can now investigate which temperature is suited to estimate the superconducting correlations. In Fig.~\ref{wixkwlqbjxaz} we 
plot the correlation length $\xi(t)$ of the Floquet system ($U=-1$, $t_\text{hop}=1/3$) as function of temperature 
(up to a rescaling, this is the same as Fig.~\ref{fig:fig1aaa}b). In addition, we extract the correlation length 
$\xi(t)$ from the driven system (Fig.~\ref{fig:fig2khqwxb}) at different times, and plot the result against the low-energy 
temperature $T_\text{FS}(t)$ (filled circles). One can see that the Cooper-pair correlations in the driven state 
follow more closely the equilibrium behavior set by the effective temperature of the electrons at the Fermi surface, while 
the correlation length at the temperature estimate from the mean energy would be considerably shorter.

Finally, we remark that similar to the electronic temperature, one can also extract a temperature of the 
bosonic fluctuations, as $T_{D}^{-1}=-\frac{d}{d\omega} [\mathcal{D}^<(\omega,t)/\mathcal{D}^>(\omega,t)]_{\omega=0}$, in 
analogy to Eq.~\eqref{khabqcblca}. This temperature is comparable in magnitude to $T_\text{FS}$ (see dashed line in Fig.~\ref{fig:fig3}b).

\begin{figure}[tbp]
\begin{center}
\includegraphics[width=0.9\columnwidth]{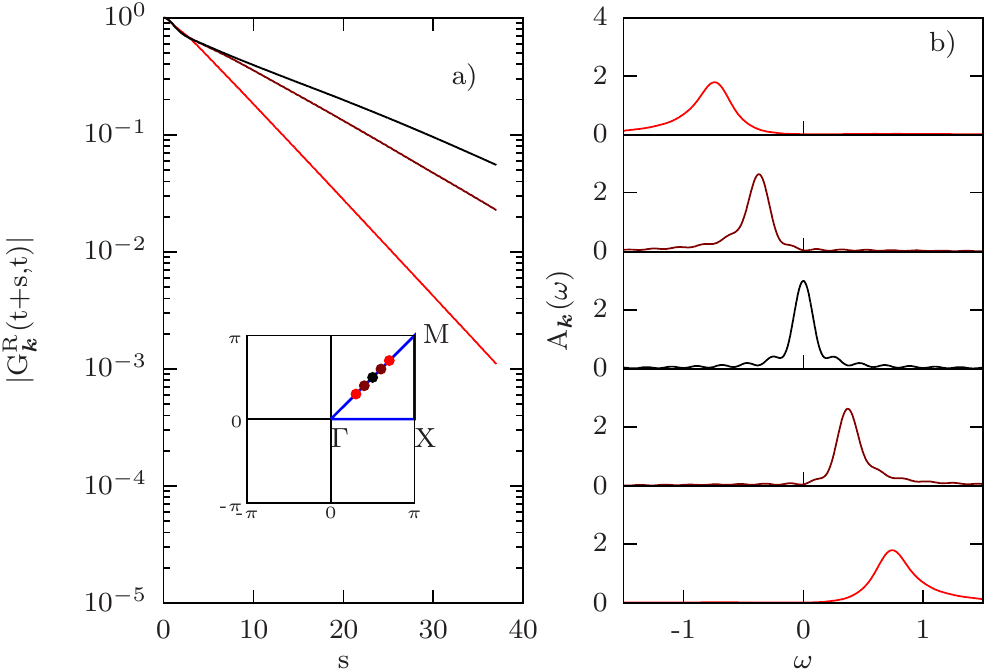}\\
\caption{(a) The modulus of the retarded Green's function at time $t$=23 for values of $\bm k$ along the $\Gamma$-$M$ 
direction of the Brillouin zone($\Omega=10.47$). Inset: The corresponding $\bm k$ points in the Brillouin zone. (b) Momentum resolved spectral function as a 
function of frequency for the same momentum.}
\label{fig:fig5}
\end{center}
\end{figure}

\subsection{Electronic spectra}
\label{hsvkqxaz04}

\begin{figure}[htb]
\begin{center}
\includegraphics[width=0.9\columnwidth]{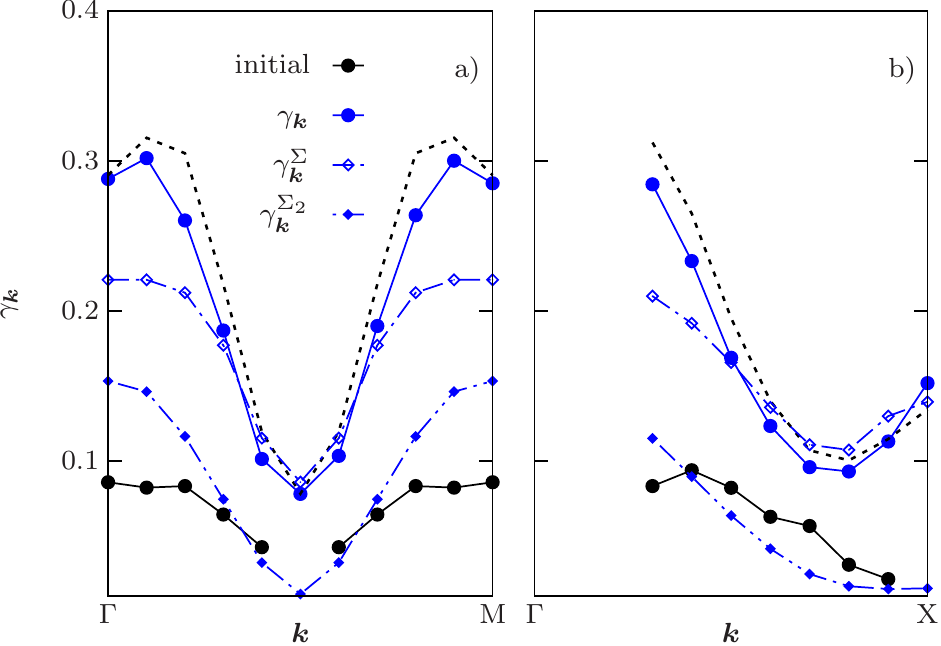}
\caption{(a) Relaxation rate  obtained from the real-time data Eq.~\eqref{eq:FL} (circles, $\gamma_{\bm k}$), from the 
self-energy [c.f.~Eq.~\eqref{pppsssps}] (open squares, $\gamma_{\bm k}^{\Sigma}$), and from the second-order 
self-energy (filled squares, $\gamma_{\bm k}^{\Sigma_2}$), for values of $\bm k$ along the $\Gamma$-$M$ direction in the 
Brillouin zone at time $t=23$ after the ramp ($\Omega=10.47$). The dashed line shows $\gamma_{\bm k}$ for the 
effective Hamiltonian ($t_{\text{hop}}$=1/3, $U=-1$) in equilibrium at 
temperature $T=T_\text{FS}=0.10$. (b) Relaxation rate $\gamma_{\bm k}$ for momenta $\bm k$ along 
the $\Gamma$-$X$ direction, $X=(0,\pi)$.}
\label{fig:fig6}
\end{center}
\end{figure}

Superconducting fluctuations in the normal state  can have a strong effect on the electronic spectra. They can give rise to a 
pseudo-gap behaviour, and for ramps close to the superconducting transition an anomalous increase of the quasi-particle lifetime 
close to the Fermi-surface has been predicted. \cite{Lemonik2017} For the parameters investigated here, there is no pseudo-gap in 
the local density of states (see Fig.~\ref{fig:fig3}a). This is not un-expected, as the effective temperature (both $T_\text{FS}$ and 
the estimate from the global energy density) is too high for a pseudo-gap to appear even in an equilibrated system. For a more detailed 
comparison of the electronic spectra of the effective Hamiltonian ($t_{\text{hop}}=1/3$ and $U=-1$) and the driven system we therefore analyze the 
momentum-dependent Green's functions $G_{\bm k}$ and the Fermi-liquid properties of the system.

In a Fermi-liquid, the momentum dependent retarded Green's function in time has the asymptotic quasi-particle form
\begin{equation}
G_{\bm k}^R(t) \sim G_{\bm k}^{coh}(t) \equiv -i Z_{\bm k} e^{-i{\tilde{\epsilon}}_{\bm k} t} e^{-\gamma_{\bm k} t}
\label{eq:FL}
\end{equation}
where ${\tilde{\epsilon}}_{\bm k}$ is the quasi-particle energy, and $\gamma_{\bm k}$ the relaxation rate. One can therefore directly 
extract $\gamma_{\bm k}$ from the real-time data: In Fig.~\ref{fig:fig5}a, we exemplarily show $|G^R_{\bm k}(t+s,t)|$ at time $t=23$ for 
values of $\bm k$ along the $\Gamma$-$M$ direction of the Brillouin zone. The functions decay exponentially at long times ($s$), from 
which $\gamma_{\bm k}$ is extracted.  This procedure is equivalent of measuring the Lorentzian line-width of the momentum-resolved 
spectra $A_{\bm k}(\omega,t)$. Just for illustration, we plot in Fig.~\ref{fig:fig5}b the spectra $A_{\bm k}(\omega,t)$, obtained from the Fourier transforms of the real time data $G^R_{\bm k}(t+s,t)$ using Eq.~\eqref{eq:FT}.

Figure \ref{fig:fig6}a shows $\gamma_{\bm k}$ along the $\Gamma$-$M$ direction of the Brillouin zone for different times. The 
electron relaxation rate is consistent with the finite-temperature Fermi liquid form
\begin{equation}
\gamma_{\bm k} \sim C{|\bm k-\bm k_F|}^2 +  C',
\end{equation}
but shows a marked increase at $\bm k=\bm k_F$ compared to the initial state. In analogy to the superconducting 
fluctuations, we find that the relaxation rates match fairly well the behavior of the effective Hamiltonian system in equilibrium at 
temperature $T_\text{FS}$ (see dashed lines). Consistent with this observation, we now show that the increase 
of $\gamma_{\bm k_F}$ with respect to the initial state can largely be assigned to the 
coupling of electrons and  Cooper-pair fluctuations.

The effect of the Cooper-pair correlations on the electronic  spectra can be quantified as follows: We first confirm 
that the lifetime of the quasiparticles can also be obtained from the self-energy. The estimate
\begin{align}
\label{pppsssps}
\gamma_{\bm k}^{\Sigma}
=
-\text{Im} \Sigma^R(\omega=\tilde \epsilon_{\bm k},t),
\end{align}
measured at some time $t$ in the driven state, accurately reproduces $\gamma_{\bm k}$ from the real-time 
data, compare circles and open squares in Fig.~\ref{fig:fig6}a. (To first approximation, $\tilde \epsilon_{\bm k}$ in Eq.~\eqref{pppsssps} is 
taken as the bare band energy $\mathcal{J}_0(A_f) \epsilon_{\bm k}=\epsilon_{\bm k}/3$, since we are anyway mainly 
interested in the properties at the Fermi surface $\epsilon_{\bm k}=0$.)  Furthermore, we can then obtain the 
contribution $\gamma_{\bm k}^{\Sigma_2}$,  by taking only the second-order self-energy $\Sigma_{\bm k}^{(2)}$ in 
Eq.~\eqref{pppsssps}, which does not take into account the Cooper-pair correlations. (To obtain $\Sigma_{\bm k}^{(2)}$, the full 
propagator $\mathcal{D}$ in Eq.~\eqref{kqcebqxl01} is replaced by Eq.~\eqref{kqcebqxl02}. Note that this is only a 
decomposition of the different contributions to $\Sigma$; the driven state is always evaluated with the full self-energy). From 
the comparison of the relaxation rates $\gamma_{\bm k}^{\Sigma_2}$ and $\gamma_{\bm k}^{\Sigma}$, we see that the increase 
of the scattering rate at the Fermi surface in the driven state can 
be attributed mainly to the interaction with the Cooper-pair fluctuations. This observation is consistent with Ref.~\onlinecite{Lemonik2017}, 
where the build-up of superconducting fluctuations lead to an anomalous peak of the scattering rate at the Fermi energy for a 
system that was quenched right to the superconducting transition. In the present case, however, the impact of the fluctuations on 
the scattering rate is rather featureless, as the driven system is at a higher effective temperature and further from a 
phase transition, so that the superconducting correlations extent only over few lattice constants.

The increase of the scattering rate in the driven state compared to the initial state is also visible along the $\Gamma-X$ direction  (Fig.~\ref{fig:fig6}b). Interestingly, in this case there is a pronounced peak in the scattering rate at $\bm k=X$. This anomalous behavior  is actually not exclusively due to the pairing fluctuations. Although small on the scale of Fig.~\ref{fig:fig6}b, it can already be seen in the scattering rate $\gamma^{\Sigma_2}_{\bm k}$ obtained from the (filled squares), and also in the $\gamma^{\Sigma_2}_{\bm k}$ in equilibrium at temperatures $T>0$ (not shown). The enhanced scattering is simply a consequence of the flat dispersion $\epsilon_{\bm k}$ at $\bm k=X$, which is the origin of the van-Hove singularity in the density of state at $\omega=0$. However, as one can see from the comparison of $\gamma^{\Sigma}_{\bm k}$ and $\gamma^{\Sigma_2}_{\bm k}$, the effect is greatly enhanced by the coupling of electrons and superconducting fluctuations. The latter suggests the interesting experimental possibility of exploiting the van-Hove points to amplify the signature of the fluctuations in the electronic spectra. In the present case, the van-Hove point accidentally lies on the Fermi-surface, but in general one can think of deforming the band structure through Floquet engineering in such a way that a van-Hove point is shifted to the Fermi surface. While this will typically require strong fields, such that an enhancement of electronic orders is possibly only transiently, our simulation suggests that at such a van-Hove point even the short range fluctuations reachable in a transient Floquet engineering protocol can become evident, making this an experimentally viable pathway.

\section{Conclusion}
 \label{hsvkqxaz001}

In conclusion, we have studied the dynamical enhancement of short-range  superconducting fluctuations 
in the  attractive Hubbard model following a renormalization of the bandwidth by a time-periodic electric 
field with frequencies close to the bandwidth. In the effective (Floquet) Hamiltonian picture, which is asymptotically 
correct in the limit of large driving frequency, the driving corresponds to a reduction of the hopping matrix elements, 
and hence an increase of the ratio interaction over bandwidth. At finite frequency, the system constantly absorbs energy from 
the drive and therefore does not reach a long-range ordered state in the long-time limit. Instead, short range correlations increase 
at short times, and  decrease as the system subsequently heats up. Based on  numerical simulations, our main observations are the 
following: (i) Even with driving frequencies close to the bandwidth, a substantial enhancement of the short-range superconducting 
fluctuations can be achieved at least transiently. (ii) The driven state at intermediate times is rather non-thermal. It cannot be 
described by an effective temperature state of the effective Hamiltonian, and a temperature estimate $T_\text{FS}$ from the 
electrons close to the Fermi surface is still lower than an estimate based on the global energy in the system. Superconducting 
fluctuations in the transient state are determined more accurately by the effective temperature $T_\text{FS}$ and are thus 
more ``robust'' against the energy absorption. (iii) The superconducting fluctuations lead to an increase of the electronic 
quasiparticle scattering rate. While for the short-ranged correlations this increase is in general rather featureless over the Brillouin zone, the  effect of short-range fluctuations is strongly enhanced at a van-Hove point in the band structure.

Although it is clearly challenging to use Floquet engineering as a way to induce long-range orders out of a gapless metallic state, our work demonstrates that the manipulation of short-range orders is in range experimentally, with 
frequencies that do not have to be far detuned from the bandwidth. Using the amplifying effect of van-Hove singularities, it may be possible to observe such a transient Floquet control of electronic orders. (Furthermore, indirect signatures of short-range superconducting correlations have been predicted in the optical conductivity.\cite{Ido2017, Bittner2017, Lemonik2018}) An interesting pathway for further investigations is also the control of (short range) charge order, which can be monitored more directly using time-resolved X-rays from free electron lasers. Charge-ordered systems (or systems where charge-order and superconductivity are intertwined) are also more strongly coupled to the lattice, which may eventually even stabilize different driven states at longer time. 

\acknowledgments
We acknowledge discussions with Denis Golez, Yuta Murakami, Philipp Werner, and Aditi Mitra. This work was supported 
by the ERC starting grant No.~716648. The calculations have been done at the RRZE of the University Erlangen-Nuremberg.

\bibliographystyle{apsrev4-1}
\bibliography{apssamp}
\end{document}